# Reversible and Persistent Photoconductivity at the NdGaO$_3$/SrTiO$_3$ Conducting Interface

*Umberto Scotti di Uccio,[*] Carmela Aruta,[*] Claudia Cantoni,[**] Emiliano Di Gennaro,[*] Alessandro Gadaleta,[***] Andrew R. Lupini,[**] Davide Maccariello,[*†] Daniele Marré,[***] Ilaria Pallecchi,[***] Domenico Paparo,[*] Paolo Perna,[*†] Muhammad Riaz,[*] and Fabio Miletto Granozio[*]*

[*]*CNR-SPIN and Dipartimento di Scienze Fisiche, Univ. di Napoli "Federico II", Compl. Univ. di Monte S. Angelo, Via Cintia, I-80126 Napoli, Italy*

[**]*Materials Science and Technology Division, Oak Ridge National Laboratory, 1 Bethel Valley Road, Oak Ridge, TN 37831-6116, USA*

[***]*CNR-SPIN and Dipartimento di Fisica, Univ. di Genova, via Dodecaneso 33, 16146, Genova, Italy*

[4]*Present address: IMDEA-Nanociencia, Madrid, Spain*

The interface between the band gap insulators LaAlO$_3$ and SrTiO$_3$ is known to host a highly mobile two-dimensional electron gas. Here we report on the fabrication and characterization of the NdGaO$_3$/SrTiO$_3$ interface, that shares with LaAlO$_3$/SrTiO$_3$ an all-perovskite structure, the insulating nature of the single building block and the polar-non polar character. Our work demonstrates that in NdGaO$_3$/SrTiO$_3$ a metallic layer of mobile electrons is formed, with properties comparable to LaAlO$_3$/SrTiO$_3$. The localization of the injected electrons at the Ti sites, within a few unit cells from the interface, was proved by Atomic-scale-resolved EELS analyses. The electric transport and photoconduction of samples were also investigated. We found that irradiation by photons below the SrTiO$_3$ gap does not increase the carrier density, but slightly enhances low temperature mobility. A giant persistent photoconductivity effect was instead observed, even under irradiation by low energy photons, in highly resistive samples fabricated at non-optimal conditions. We discuss the results in the light of different mechanisms proposed for the two-dimensional electron gas formation. Both the ordinary and the persistent photoconductivity in these systems are addressed and analyzed.

The search for novel oxide heterostructures supporting the formation of a two dimensional electron gas (2DEG) at the interface between two robust band insulators is of major interest both for fundamental and applied physics.[1] The availability of multiple systems implementing such physical concept would offer in fact some important advantages: increasing, on the one hand, the degrees of freedom for scientists trying to address the physical mechanisms and the material issues underlying 2DEG formation; opening the route, on the other hand, to optimizing device properties by adopting suitable materials in view of specific applications.

In a recent paper, the conducting properties of the LaGaO$_3$/SrTiO$_3$ (LGO/STO) interface where reported.[2] The NdGaO$_3$/SrTiO$_3$ (NGO/STO) epitaxial heterostructure addressed in this work shares with both LaAlO$_3$ (LAO)/STO,[3,4,5] and LGO/STO structures,[2,6,7] some features considered crucial for electronic reconstruction (ER) to take place: i) NGO is constituted by a stack of planes with alternate charge, thus creating a polar discontinuity when epitaxially grown on STO; ii) NGO possesses a larger gap (3.8 eV) than STO. This latter feature has been proposed as an important indicator in polar-non polar interfaces,[2] because it allows the conduction band of the polar layer to be presumably above the conduction band of STO, even when considering relative band alignment. Interestingly, NdGaO$_3$ doesn't contain La, which was suspected to dope the interface and to play an important role in the origin of LAO/STO conductivity.[8,9,10]

The transport properties of LAO/STO heterostructures are sensitive to light irradiation. This effect was first considered as detrimental for the realization of correct measurements of the intrinsic transport properties.[11,12,13] More recently, the photoconductivity of LAO/STO has attracted a considerable interest in view of potential optoelectronics applications and as a possible tool to investigate the fundamental physics of the system.[14,15,16] Contrasting evidences have been reported so far about the threshold photon energy of LAO/STO photoconductance: room temperature measurements performed on metallic samples (thickness of LAO above 4 unit cell) showed a threshold for photoconductivity at about 380 nm,[12] corresponding to the indirect $SrTiO_3$ gap (about 3.3 eV);[17,18] sensitivity to visible light (i.e., to photons with subgap energy) was instead reported for insulating, 3 unit cell thick LAO;[11,14] finally, annealed samples with high resistance and non metallic behavior showed 5 orders of magnitude drop of resistance under irradiation at 395 nm (about 3.1 eV, slightly below the indirect gap threshold).[16]

Here we show that NGO/STO interfaces host a conductive 2DEG, the transport properties of which are affected by irradiation with visible and UV light. Our data prove that the photoresponse of samples is strongly dependent on their electrical resistance at room temperature: metallic samples are only marginally affected by subgap photons; highly resistive samples are instead effectively photo-doped by subgap photon irradiation and show giant, persistent photoconductivity.[16] The overall information provided by our data is then analyzed, with the aim of understanding to what extent NGO/STO shares with LAO/STO analogous fundamental mechanisms.

**EXPERIMENTAL RESULTS**

Films of NGO (and of LAO, LGO for comparison) were deposited on nominally $TiO_2$ terminated STO substrates, chemically treated in de-ionized water and buffered-HF. During NGO deposition, the RHEED intensity oscillations (see the section devoted to Methods for more details) indicate layer-by-layer growth (Figure 1a). After a transient, the oscillations are regular; the magnification (Figure 1b) also demonstrates the effect of each laser shot. In Figure 1c we show a typical RHEED pattern after 10 unit cells deposition. The half order rods are reminiscent of the surface of NGO single crystals, and were associated by LEED to a c(2×2) reconstruction,[19] in agreement with the orthorhombic distortion of the pseudocubic NGO unit cell. The surface reconstruction persists until the end of the deposition, with a 2D pattern demonstrating a high crystallinity.

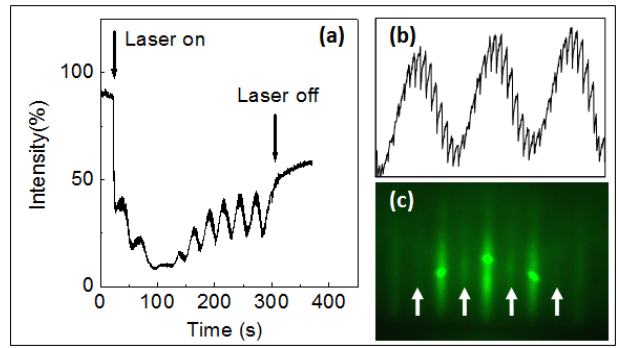

**Figure 1.** a) RHEED oscillations during the growth of a NGO film; b) magnification of a part of Figure 1a; c) final RHEED pattern. Arrows indicate the fractional reflections associated with the 2×2 reconstruction.

The atomic structure of two different electrically conducting NGO/STO samples deposited at 800°C was investigated by aberration corrected scanning transmission electron microscopy (STEM) and electron energy loss spectroscopy (EELS) measurements. The images revealed that the NGO films were free of structural defects and had grown with a cube-on-cube epitaxial relationship of their pseudocubic cell on the STO cell (Figure 2). The HAADF intensity and EELS spectra across the interface revealed a stacking sequence of the type $SrO-TiO_2-NdO-GaO_2$ as indicated in the inset model structure of Figure 2c. Although the interface appeared not atomically abrupt, the observed cation intermixing was limited to 1-2 unit cells. The spatial distribution of electronic charge present at the STO interface was derived by analyzing changes in the fine structure of the EELS $Ti-L_{2,3}$ and O-K edges associated to a change of the Ti valence as function of depth. Shown in Figure 2a are some of the O-K edges extracted from an EEL spectrum image across the interface. The data demonstrate that the extra electrons are hosted by the $t_{2g}$ levels associated to the Ti 3d states. The Ti valence is proportional to the kinetic energy difference ΔE between peaks A and C in the O-K edge.[20] The depth profile for the Ti valence (Fig. 2b) follows an exponential law, indicating a confinement depth of ~ 3 unit cells. The deduced total injected charge (0.8 ± 0.2 e/unit cell area) is overestimated; this is due to oversampling of O atoms in the NGO environment caused by cation intermixing, which generates a shoulder in peak C on the STO side (see dashed line in Fig. 2a), thus complicating the quantitative analysis.

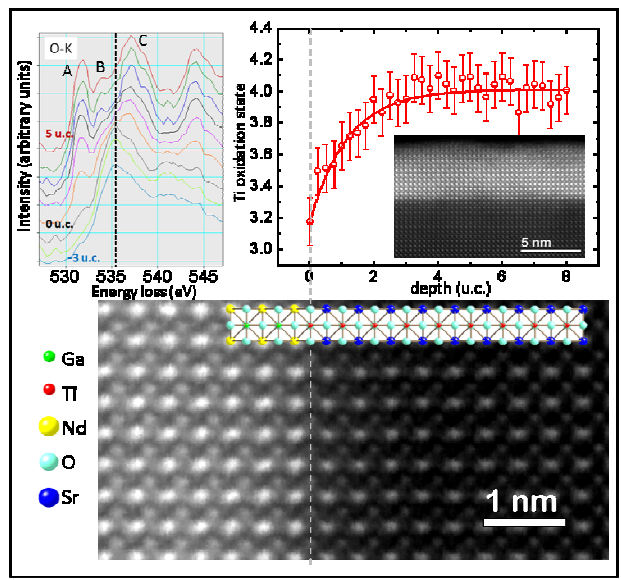

**Figure 2.** a) O-K edges extracted from an EEL spectrum image across the interface and shown at spacing of 1 unit cell. b) Ti valence depth profile derived from the analysis of the O-K fine structure in the same spectrum image, and HAADF image of the NGO/STO interface (inset). c) Higher magnification HAADF image of the interface with corresponding sketch of atomic stacking and same scale as the Ti valence plot.

The R(T) curves of NGO/STO are qualitatively similar to those of LAO/STO and LGO/STO (Figure 3). In all three cases a metallic behavior is observed in a wide range of temperature and, in most samples, at any temperature. The slight resistance upturn, which in LAO/STO is ascribed to weak carrier localization or possibly to the Kondo effect,[21,22] is sometimes observed below ≈50K in NGO/STO as well (Figure 3, right panel).

The following procedure was adopted to investigate the photoconductivity. Fresh samples were kept in dark for several days, and a first transport characterization was performed. During the second run of measurements, the samples were exposed to the light produced by different sources in the range visible to UV. In order to exclude trivial effects of above-gap excitation in the substrate, a series of test measurements on bare STO were made, confirming that photoconductivity in those crystals is immeasurably low in our measuring configuration and under every light source used in this work. The sheet resistance of a NGO/STO sample in dark and in presence of a 402 nm radiation are compared in Figure 3 (right panel). The measurement shows that subgap excitation does not affect the resistance at high temperature, as also previously reported for LAO/STO;[12] however, a significant reduction of resistance is observed in the lowest temperature region. In order to clarify this behavior, we fitted the curves by the phenomenological expression recently proposed to describe the behavior of R(T) in the 2DEG induced at the STO surface by field effect,[22]

$$R = R_o + aT^2 + b\,T^5 + c\left(\frac{T_o^2}{T^2 + T_o^2}\right)^\alpha$$

. The four terms describe the residual scattering, the electron-electron and electron-phonon interaction, and the Kondo effect respectively. Following the mentioned paper, the parameter $\alpha$=0.225 was kept constant. The best fit values for the 5 parameters share the values of a and b, while the Kondo-related term disappears under irradiation. All values are reported in tab. 1. In other terms, the Kondo term is completely suppressed by light, and an increase of residual resistance takes place. Further insight is achieved by carrier density and Hall mobility measurements performed in dark and under 402 nm UV irradiation. The results for the same sample of Figure 3 (right panel) are reported in Figure 4a,b. In dark, the sheet carrier density is ≈3×10$^{14}$ cm$^{-2}$ at room temperature, and the mobility is ≈1 cm$^2$V$^{-1}$s$^{-1}$. In the higher temperature region (>100K) the mobility increases with decreasing temperature following a T$^{-2}$ law, with low temperature saturation at values up to 10-100 cm$^2$ s$^{-1}$ V$^{-1}$, which is somewhat lower than typical low temperature values reported for optimized LAO/STO samples.[23] This result, together with R(T) fits reported above, shows that the influence of subgap irradiation on transport doesn't stem from a variation in the number of free carriers, but instead from an enhancement of their low temperature mobility. Conversely, when samples are irradiated with the Hg lamp, which emits a number of spectral lines well above gap, the photo-conductance effect is significant at any measured temperature. The relative resistance reduction is of the order of about 20-40% at room temperature, in agreement with previous reports for LAO/STO,[15] increasing to a factor 2-5 at 10K. This behavior is consistent with the idea that above-gap photons efficiently promote electrons from the valence to the conduction band. This effect is also persistent: after turning the light off, the characteristic times of recovery may reach up to thousands of seconds.

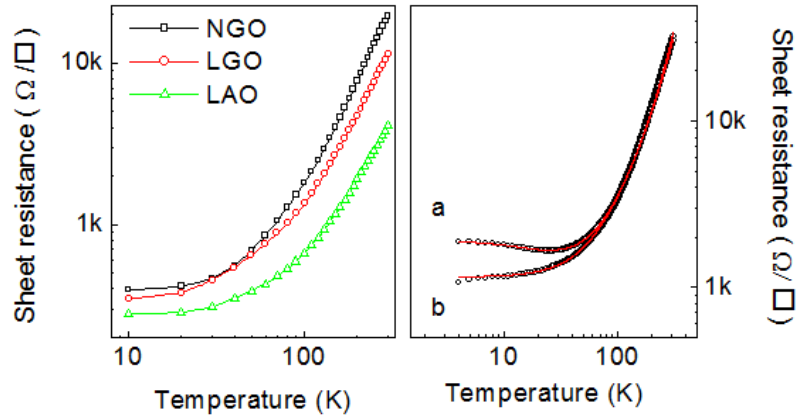

34

**Figure 3.** Left panel: sheet resistance of NGO/STO, LGO/STO, LAO/STO in dark. Right panel sheet resistance of NGO/STO; a) in dark; b) under UV (402 nm) irradiation. Solid curves are fits.

| | a ($\Omega K^{-2}$) | b ($\Omega K^{-5}$) | $R_o$ ($\Omega$) | c ($\Omega$) | $T_o$ (K) |
|---|---|---|---|---|---|
| *Under irradiation* | 0.23 | $3\times10^{-9}$ | 700 | 1200 | 12 |
| *In dark* | | | 1150 | 0 | - |

**Tab. 1** Best fit values derived for the two R(T) curves in Figure 3b

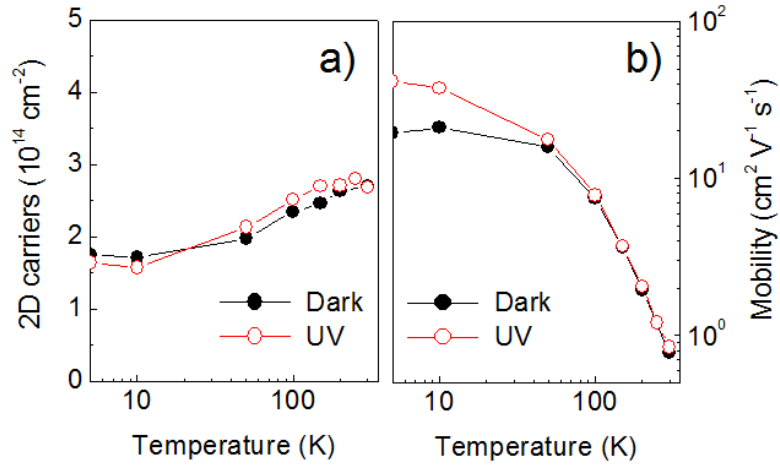

**Fig.4 a)** Sheet carrier density and b) mobility vs. temperature for the same NGO/STO sample as in Fig. 3b

NGO/STO samples fabricated at high temperature were systematically less stable than samples grown at 700°C (which we chose as a benchmark) and they often showed high room-temperature resistance, non-metallic behavior and a much stronger photoresponse. A similar but much weaker dependence of photoresponse intensity on deposition temperature was also reported for LAO/STO.[15] EELS measurements suggest that this effect may be related to a higher reactivity of NGO with STO (e.g., when compared with LAO), resulting in some blurring of the interface in samples grown at $T_s \approx$ 800 °C. In the following, we report on the transport properties of such samples.

Non metallic NGO/STO showed a giant persistent photoconductivity at any measured temperature, both in the case of subgap irradiation and when visible light is employed. Figure 5 shows the response of an insulating sample, which is driven to the metallic state by irradiation with different light sources. Even the radiation of a bulb lamp dramatically reduced the resistance, with a decrement of orders of magnitude at low temperature. However, for equal exposure times, the lower is the wavelength, the larger the effect. The best metallic behavior was achieved when the radiating spectrum contained a fraction of above-gap photons (i.e., when a UV Hg lamp was used). Qualitatively, the data can only be understood by assuming that even subgap photons effectively promote charge carriers to the conduction band. The mobility is comparable to that of conductive samples; as an instance, a formerly insulating sample which showed under irradiation a carrier density of $4\times10^{12}$ cm$^{-2}$ carrier density had 0.2 cm$^2$ V$^{-1}$s$^{-1}$ mobility at 300 K.

The photo-response of the same sample was recorded vs. time by the following procedure. The sample was exposed for a long time to a UV (420 nm) radiation, which was briefly turned off. When the radiation is turned on again, the conductance begins to increase, approaching its original value with a rate that gets smaller with time. When the light is turned off again, the conductance decreases showing a similar dependence on time. The slow variations are characterized by a scaling behavior, which is universal in temperature. In order to quantitatively analyze the data, the function

$$\psi(t) = \frac{\sigma(t) - \sigma_o}{\sigma(t_1) - \sigma_o}$$

(being $\sigma_o$ the initial conductivity; $t_1$ an arbitrary final time and $\sigma(t_1)$ the corresponding value of conductance) is plotted vs. time in Figure 6. This scaling law is easily understood in terms of the Drude model, by assuming that the enhancement of conductivity $\Delta\sigma(t)$ is given by $\Delta\sigma(t, T) = \Delta n(t) \dfrac{e^2 \, \tau(T)}{m}$. This is different from the universal scaling of the metal insulator transition (MIT) in, e.g., semiconductors, where the dependences of $\sigma$ on $\Delta n$, T are entangled according to $\sigma(\Delta n, T) \propto \exp\left( A\,\Delta n \Big/ T^{\frac{1}{z\nu}} \right)$,[24] but consistent with MIT investigations in LAO/STO.[25]

Besides, the scaling rules out phonon-assisted mechanisms in the optical transitions related to the photoresponse. Either stretched exponential or logarithmic functions were employed to fit the temporal evolution of photoresponse.[15,26,27] In the range of applicability, both expressions provide a satisfactory fit. The solid curve in fig.6 is the plot of the function

$$\psi(t) = A\left( 1 - \exp\left( -\left(\frac{t}{t_o}\right)^{\beta} \right) \right)$$

, with $t_o$=17s and $\beta$=0.64.

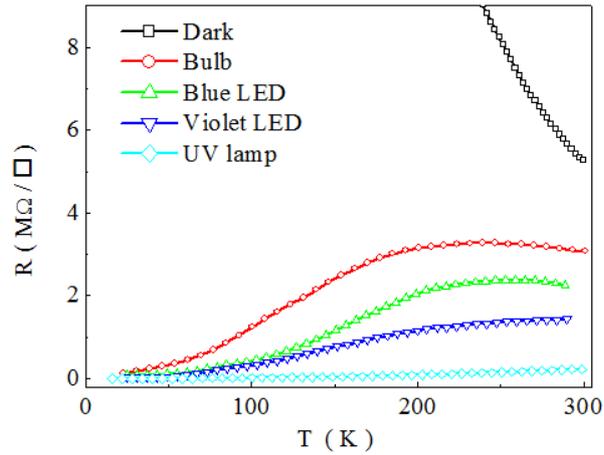

**Figure 5.** Photoconductivity of a highly resistive NGO/STO sample.

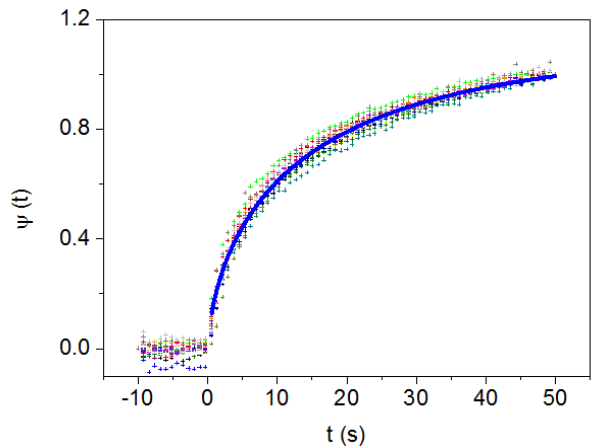

**Figure 6.** Normalized photoresponse vs. time (UV LED at 420 nm). Symbols are data taken at 20 different temperatures in the range 100-300K. The solid curve is a fit (see text).

The previously described procedure does not test the time needed to recover the original state after light is turned off. This is extremely long, as shown in Figure 7. The black curve was collected before any intentional irradiation and keeping the sample for several days in dark after exposure to natural light. The sample was then turned metallic by UV radiation (Hg lamp, bottom blue curve). After turning the radiation off, only a partial recovery took place. The green and red curves where recorded after keeping the sample in darkness for about 2 and 4 months, respectively. After several months, the room temperature resistance was still ≈3 times smaller than for the same sample when firstly irradiated. We stress that this persistent behavior is observed whenever the photoresponse is dominated by photo-generation of new carriers: i.e., it also takes place in low resistance samples when irradiated by above-gap photons, even though to a less dramatic extent.

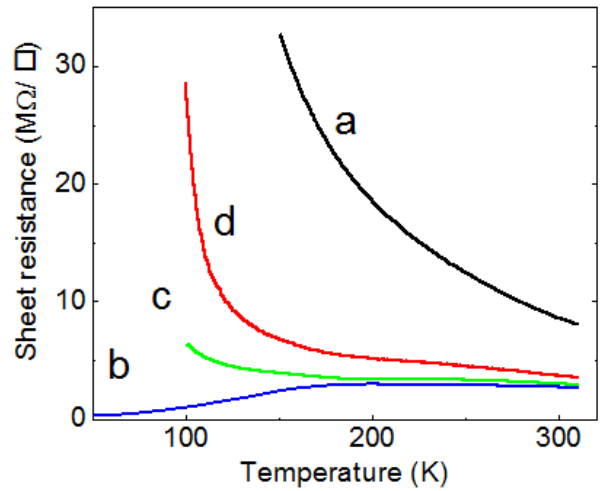

**Figure 7.** a) R(T) of at a highly resistive sample previous to exposure to UV light; b) immediately after exposure; c, d) after subsequent exposure to darkness for 2 and 4 months, respectively.

**DISCUSSION**

By comparing our data with literature, we deduce a scenario where the electrical transport and the nature of the photoresponse are both determined by the room temperature resistance values of the polar-non polar heterostructure (whether it is LAO/STO, or NGO/STO) rather than by the chemistry of the polar material. This statement is based on three main observations, as described in the following.

i) LAO/STO interfaces with standard room temperature resistance values of the order of 10kΩ are unperturbed at room temperature by photons below the STO indirect gap, and are slightly perturbed, with moderate R decrease (20-40%) by radiation above the indirect gap.[12,15] We found that this is also true for NGO/STO with comparable room temperature resistance. We demonstrated that subgap irradiation doesn't promote new carriers in the conduction band of such interfaces. The moderate increase of conductivity at low temperature is due to the suppression of localization phenomena, which are well described by assuming the existence of Kondo scattering.

ii) highly resistive LAO/STO heterostructures ($R \approx 10^6 \Omega$ or above) are sensitive to subgap irradiation,[16] as well as (insulating) standard 3 unit cells thick LAO/STO samples.[11,14] This is also true for highly resistive NGO/STO.

iii) when irradiated by UV light, both highly resistive samples of LAO/STO,[28] and of NGO/STO, undergo reduction of resistance persisting on extremely long time scales.

The above reported considerations lead to two crucial questions i): is there enough evidence, in view of the similarities of transport properties, to state that the conductive behavior in NGO/STO has the same origin as in LAO/STO and LGO/STO? ii) What is the microscopic mechanism determining the effect of light on NGO/STO heterostructures? We try to answer in the following, by separately addressing the relevant issues.

**Possible mechanisms of conductivity**

The EELS measurements show the existence of a very thin region of STO, confined within about 3 unit cells from the interface, where the Ti valence is intermediate between 3+ and 4+, where the Ti 3d states forming the STO conduction band are therefore partially populated. Our data then provide a direct evidence of the nm-scale confinement of the free electrons that are responsible for the interface conductivity, i.e., of its 2D nature, similarly to what observed in the LAO/STO interfaces and predicted by theoretical models of that system.

An exhaustive discussion of the possible mechanisms that make NGO/STO interfaces conducting is beyond the scope of this paper. Nevertheless, some considerations that emerge from our work can possibly contribute to clarifying this issue. First of all, NGO/STO possesses all the physical characteristics considered a prerequisite to the ER, which results in the injection of carriers from the polar layer into STO, where they finally reside in a quantum well close to the interface. EELS measurements, in spite of the observed level of interdiffusion, are also fully compatible with the ER scenario. At the same time, different mechanisms that were proposed for LAO/STO are also possible for NGO/STO and cannot be completely ruled out by those measurements. Among the others, we comment here on cation intermixing and on oxygen vacancy formation.

Cation intermixing at the LAO/STO interface due to La subplantation was indicated as a source of extrinsic doping that can produce free carriers. We mention that the dynamics of Nd and La during the PLD are quite the same, because the atoms have very similar mass. A recent study demonstrated that the kinetic energy of such species is not sufficient to allow implantation across several unit cells.[6] However, we don't exclude a higher reactivity of Nd with STO at high temperature (i.e., 750-800°C), which might explain the mentioned sensitivity to deposition conditions and the results of EELS measurements. Following this argument, one is driven to the conclusion that intermixing just worsen the properties of non-optimized NGO/STO; that is, the occasional high resistance may originate from disorder at the interface. A deeper microstructural/ microelemental investigation is required to be conclusive on this point.

Doping of STO by oxygen vacancies is another possible extrinsic source of free charges in STO-based interfaces, and one might wonder about their role in NGO/STO. However, we obtained conductive samples up to $5 \times 10^{-2}$ mbar oxygen pressure during deposition. Such value is considered high enough to exclude that oxygen vacancies determine the transport properties of STO/LAO.[23, 29] The oxygen vacancies content may be even lower in the case of NGO/STO, in view of the smaller affinity for oxygen of Ga with respect to Al.[30] Also note that during a number of tests of high temperature electric transport some samples were heated up to 450°C in air, a condition in which oxygen vacancy doped STO is known to fully recover its insulating properties. Nevertheless no measurable hysteretic effect in the warming-cooling cycle was found. Measuring at higher temperatures was precluded by contacts failure.

**Possible mechanisms for reversible and persistent photoconductivity**

Before discussing the properties of interfaces, let's address the behavior of bare STO. Even under irradiation by the above-gap lines of our Hg lamp, no photoresponse was detected within the sensitivity of our instruments. This can be understood on the basis of some quantitative considerations in terms of the short recombination time of excitons in bulk STO (a few ns).[31] By assuming a photon energy of about 4 eV, a recombination time of about 2-3ns and an exciton generation efficiency of 100% (upper limit), an incident above-gap UV component exceeding $10^{20}$ photons cm$^{-2}$s$^{-1}$ (i.e.> 10W/cm$^2$ , orders of magnitude above the fluence of our Hg lamp) would be needed to sustain a surface electron density of $10^{12}$ e/cm$^{-2}$, suggested to be necessary to induce an insulator to metal transition in STO.[32] Once spurious contributions from bulk STO are excluded, all the observed effects must be explained in terms of interface properties. Since the only physical parameter contained in the above considerations that can be affected by the presence of the interface is the exciton lifetime, our argument brings us to the conclusion that the interface photoconductivity is related to a giant enhancement of the photo-carrier lifetime with respect to bare STO.

We discuss now the mechanism of photo-carrier generation. Starting with a very general statement, it is evident that photons can promote carriers from (more) localized to (more) mobile states in oxide polar-non polar interfaces. The final state is relatively well defined, as being in the conduction band of STO at the interface. Conversely, the variety of different responses to light reported in this work doesn't allow assuming that electrons, independently from the of radiation wavelength and from the initial sample properties, always originate from the same states. High energy photons always cause significant conductivity enhancement, which must be interpreted in terms of an effective and persistent promotion of charge carriers from the valence to the conduction band. Instead, two distinct mechanisms must be assumed for low and high resistance interfaces under irradiation of sub-gap photons.

In low resistance samples, subgap irradiation doesn't create new carriers, but just enhances low-temperature mobility of samples showing a resistance upturn at low temperature. We deduce that the electrons promoted by light undergo a transition from weakly localized initial states (where Kondo scattering is effective) to more delocalized states of STO, both belonging to the conduction band. In detail, our results are compatible with the following scenario. As known, the presence of the interface,[33] and in particular of the associated electric field,**Errore. Il segnalibro non è definito.** partially removes the degeneration of Ti 3d $t_{2g}$ states, both by splitting the energy of eigenstates with in-plane ($d_{xy}$) and out-of-plane ($d_{xz}$, $d_{yz}$) component, and by shifting in energy the states of Ti atoms at different distance form the interface. Therefore, different (and partially overlapping) sub-bands concur to the DOS of free carriers. In particular, electrons populating the states of Ti planes that are closest to the interface are susceptible to localization due to the high effective mass and strong 2D character. Such localized electrons are Ti($d^1$) centers, and have been suggested to be active Kondo scattering centers.[34] Electrons that occupy states of Ti atoms farther from the interface have smaller mass and extend over a wider region, allowing better conduction. Phenomenologically, this mechanism was also described in terms of two parallel conduction channels.[34] Thus the photo-excitation of electrons from the low-energy, localized states, to the higher energy states, both promotes conduction and suppresses Kondo scattering. Note that the intra-band optical transitions, which violate the dipole selection rules in the bulk, are instead permitted in the interface region, where the symmetry along z is broken and atomic orbitals are severely distorted by the intense electric field. Inter-site transitions (that are symmetry-allowed), either between d-states of neighboring Ti planes, and between O2p/Ti3d orbitals (that are partially hybridized) may also play a role.

High resistance samples have striking different behavior. The huge drop of resistance implies that in these samples subgap photons efficiently promote electrons from deeper, strongly localized states to the conduction band. The effect of light is therefore a typical photo-doping.

**General considerations on NGO/STO interfaces**

The overall evidence resulting from our work suggests that the presence of NGO induces deep modifications in the interfacial STO layer with respect to the free surface of STO.

i)   The Ti 3d band of high quality samples is populated by mobile electrons. In the ER scenario, the donor states are in the portion of the VB of NGO that the built-in potential lifted above the Fermi level. Furthermore, trap states located at crystal defects may also act as donors.

ii)  The optical gap of best samples is consistent with that of metallic LAO/STO. In highly resistive samples, instead, the optical gap is (at least partially) closed, presumably due to localized defect states (absent in the highly conductive samples) that can be excited by low energy photons, and resulting in the promotion of a substantial number of new carriers in the conduction band by low energy photons.

iii) The photoelectron-hole pairs lifetime is increased; in highly resistive (presumably disordered) samples even by 10 orders of magnitude. The mechanisms proposed to explain such huge increase of recombination time are usually based on the separation in space of the electron-hole pair due to local electric field.[27,35,36] A strong electric field is actually supposed to be present at polar-non polar interfaces, as suggested both by ab-initio computations,[37] and by mapping the polarization profile across the interface.[20] The observed confinement of electrons within STO at the interface of NGO/STO suggests to extend the same considerations also to this case.

Finally, we would like to remark that understanding the persistent photoconductance process basically corresponds to finding out the nature and final location of the photo-generated holes. i.e., whether the trap states are located in STO (chemical doping scenario, related to oxygen vacancies or cation interdiffusion) or in the polar layer. The former interpretation is deduced from the behavior of bulk STO.[16,17] However, since the valence band of the polar layer is bent upwards by its own built-in potential,[38] these states may as well host the traps in the quasi-equilibrium state.

In consideration of the phenomenology provided by our work, we believe that the photo-generation of carriers in insulating or highly resistive samples needs further investigation to be fully clarified.

We also consider it a very interesting subject, since the mechanism of photo-generation appears at tightly entangled to the still hot issue of the origin of the 2DEG at the polar- non polar interfaces.

**CONCLUSIONS**

Based on the similarities between LAO and NGO perovskites, we evaluated NGO/STO as a possible variant of the well known conducting LAO/STO interfaces. Samples grown by RHEED-monitored PLD were investigated by HRTEM, which showed the sequence of planes required for a polar-non polar interface. EELS analyses confirmed the electronic charge confinement at the interface in a region of a few unit cells of STO. Transport and Hall measurements demonstrated overall properties quite similar to LAO/STO, both in terms of carrier density and of electron mobility. Light irradiation also determined analogous effects: the photoresponse was determined by the room temperature resistance values of the polar-non polar heterostructure rather than by the chemistry of the polar material. In low resistance samples, subgap irradiation doesn't create new carriers, but just enhances low-temperature mobility and inhibits the resistance upturn, which we tentatively attributed to Kondo scattering. In high resistance samples, subgap photons efficiently promote electrons into the conduction band, determining a typical photo-doping effect. In the latter case, the photoconductivity has a persistent character. The huge recovery time is explained in terms of a spatial separation between excited electrons and holes and to the presence of strong interface electric fields.

**METHODS**

**Materials**. The growth was performed by reflection high energy electron diffraction (RHEED) assisted pulsed laser deposition (KrF excimer laser, 248 nm) with fluence 1.5–2.5 J cm$^{-2}$ at the target, substrate temperature $T_s$ between 700 and 800 °C and different oxygen pressures within the $5\times10^{-2}$–$10^{-4}$ mbar range. Details on the growth mode, structure and transport properties of LAO- and of LGO–based interfaces were reported elsewhere.[2,6,7,39]

**Microstructural characterization**. electron microscopy (STEM) and electron energy loss spectroscopy (EELS) measurements were performed in a Nion UltraSTEM operated at 100 kV and equipped with a third generation C3/C5 aberration corrector and an Enfina EEL spectrometer. Cross section images of the samples were acquired at several locations using both bright field (BF) and high-angle annular dark field (HAADF) detectors.

**Transport measurements.** Transport measurements, including sheet resistance vs. temperature R(T), magnetoresistance, Hall effect, and Hall mobility, were carried out on a batch of samples from 2.5 K to room temperature and in magnetic fields up to 9 T. The measurements were performed in a standard Hall bar geometry resorting to ultrasonic bonding of the contacts.

**Photoconductance measurements.** Different sources were employed: several LEDs (475 nm, 410 nm and 402 nm); an incandescent light bulb; and a Hg lamp. Assuming a black body radiation of about 3000 K, the incandescent bulb emits about 1% of its total power below 500 nm and about 0,2% below 400 nm. As known, the emission spectrum of Hg contains several intense lines in the UV region, above the gap threshold E=3.3 eV.

AUTHOR INFORMATION

**Corresponding Author**

Prof. U. Scotti di Uccio , address correspondence to scotti@na.infn.it

**Present Addresses**

† IMDEA-Nanociencia, Madrid, Spain

ACKNOWLEDGEMENTS

Financial support by EU under the project OXIDES, by MIUR under Grant Agreement PRIN 2008 - 2DEG FOXI , by European Union Seventh Framework Program (FP7/2007-2013) under grant agreement N. 264098 - MAMA, and by Compagnia di San Paolo is acknowledged. CC and ARL acknowledge funding by the US Department of Energy, Office of Science, Materials Sciences and Engineering Division.